\documentclass[preprints,article,accept,moreauthors,pdftex]{mdpi}
\usepackage[final]{changes}
\usepackage{longtable}
\firstpage{1}
\pubvolume{xx}
\issuenum{1}
\articlenumber{5}
\pubyear{2020}
\copyrightyear{2020}
\history{Received: date; Accepted: date; Published: date}
\Title{Single-element dual-interferometer for precision inertial sensing}
\usepackage[subrefformat=parens]{subcaption}
\usepackage{comment}
\usepackage{graphicx}
\usepackage[most]{tcolorbox}
\usepackage{tikz}
\usepackage{upgreek}

\newcommand{\bhline}[1]{\noalign{\hrule height #1}}

\Author{Yichao Yang$^{1,2,3,\dagger}$\orcidA{}, Kohei Yamamoto$^{1,\dagger}$\orcidB{}, Victor Huarcaya$^{1}$\orcidD{}, Christoph Vorndamme$^{1}$\orcidE{}, Daniel Penkert$^{1}$\orcidF{}, Germ\'an Fern\'andez Barranco$^{1}$\orcidG{}, Thomas S Schwarze$^{1}$\orcidH{}, Moritz Mehmet$^{1}$\orcidJ{}, Juan Jose Esteban Delgado$^{1}$\orcidI{}, Jianjun Jia$^{2,3}$, Gerhard Heinzel$^{1}$\orcidK{} and Miguel Dovale \'Alvarez$^{1}$\orcidC{}*}
\AuthorNames{Yichao Yang, Kohei Yamamoto, Victor Huarcaya, Christoph Vorndamme, Daniel Penkert, Germ\'an Fern\'andez Barranco, Thomas S Schwarze, Juan Jose Esteban Delgado, Moritz Mehmet, Jianjun Jia, Gerhard Heinzel, Miguel Dovale \'Alvarez}
\address{%
$^{1}$ \quad Max-Planck-Institut f\"ur Gravitationsphysik (Albert-Einstein-Institut) \\ and Institut f\"ur Gravitationsphysik, Leibniz Universit\"at Hannover,\\
Callinstrasse 38, D-30167 Hannover, Germany\\
$^{2}$ \quad Key Laboratory of Space Active Opto-electronics Technology, \\ Shanghai Institute of Technical Physics, Chinese Academy of Sciences, Shanghai 200083, China \\
$^{3}$ \quad University of Chinese Academy of Sciences, Beijing 100049, China}
\corres{Correspondence: \href{mailto:miguel.dovale@aei.mpg.de}{miguel.dovale@aei.mpg.de}}
\firstnote{These authors contributed equally to this work.}
\abstract{\added{Tracking moving masses in several degrees of freedom with high precision and large dynamic range is a central aspect in many current and future gravitational physics experiments. Laser interferometers have been established as one of the tools of choice for such measurement schemes.} Using sinusoidal phase modulation homodyne interferometry allows a drastic reduction of the complexity of the optical setup, a key limitation of multi-channel interferometry. By shifting the complexity of the setup to the signal processing stage, these methods enable devices with a size and weight not feasible using conventional techniques. In this paper we present the design of a novel sensor topology based on deep frequency modulation interferometry: the self-referenced single-element dual-interferometer (SEDI) inertial sensor, which takes simplification one step further by accommodating two interferometers in one optic. \added{Using a combination of computer models and analytical methods} we show that an inertial sensor with sub-picometer precision for frequencies above 10\,mHz, in a package of a few cubic inches, seems feasible with our approach. Moreover we show that by combining two of these devices it is possible to reach sub-picometer precision down to 2\,mHz. In combination with the given compactness, this makes the SEDI sensor a promising approach for applications in high precision inertial sensing for both next-generation space-based gravity missions employing drag-free control, and ground-based experiments employing inertial isolation systems with optical readout.}

\begin{document}

\section{Introduction}
\label{section:introduction}
Precision interferometry with a dynamic range over multiple fringes is the core metrology technique employed in space-based detectors, such as LISA~\cite{LISA2017}, to measure the displacements between the space-craft and free-floating test masses, or in ground-based detectors, such as LIGO~\cite{Collaboration2015}, to measure ground motion in order to isolate the suspended test masses from vibration. The pinnacle of this technology is applying such measurement to all degrees of freedom (DOF) of one or multiple test masses~\cite{Isleif2016} whilst providing increased sensitivity over other schemes such as electrostatic readout~\cite{Touboul1999} or optical levers~\cite{Acernese2004, Huarcaya2019}. Such ambitious goals require a drastic reduction of the size and complexity of the optical setup.

Standard two-beam interferometers have an operating range that is typically less than a quarter of a wavelength of pathlength difference. To increase the sensing range over many fringes, several techniques exist, such as homodyne quadrature interferometry~\cite{Cooper2018} or heterodyne interferometry~\cite{Heinzel2003}. However, these techniques suffer from some drawbacks~\cite{Watchi2018}. \added{Homodyne quadrature interferometers operate by producing two signals in quadrature (i.e., with a $\pi/2$ phase shift), from which the phase can be extracted in post-processing. They typically rely on using polarization optics and several beamsplitters, which can make it very challenging to deal with stray light noise, and sometimes employ complex actuation mechanism to reduce spurious effects in the optics.} Heterodyne interferometry does not scale well: it is too complex and bulky to be adapted to multi-DOF sensing of multiple test masses. In recent years a new interferometry technique has been developed at the Albert-Einstein-Institut: Deep Frequency Modulation (DFM) interferometry~\cite{Heinzel2010, Gerberding2015}, which combines large dynamic range and optical minimalism.

In DFM, a strong frequency modulation is applied to the laser input beam in an unequal arm-length interferometer. This modulation is embedded in the output signal in the form of components at multiple higher harmonics of the modulation frequency. The signal phase reallocates the complex amplitudes of these components in a predictable manner, which allows the extraction of the phase in real time or in post-processing by fitting the complex amplitudes of the modulation harmonics using, e.g., a Levenberg-Marquardt (least-squares) routine. This scheme requires fewer optical components, allowing for compact optical layouts or even single-element interferometers~\cite{Isleif2019}.

As with any interferometer with arms of unequal length, laser frequency noise is converted into phase (and displacement) readout noise. \added{Typically, in order to reach sub-picometer precision, interferometric inertial sensors rely on using pre-stabilized lasers, which adds to their complexity and cost. Previous DFM-based inertial sensors~\cite{Isleif2019} made use of an auxiliary interferometer~\cite{Gerberding2017} for laser frequency noise suppression, and demonstrated that it was possible to do so by performing the subtraction of the reference interferometer signal, scaled by the ratio between the interferometers' effective modulation indices, from the measurement signal, either in real time or in post-processing. In this paper we explore the possibility of accommodating the inertial sensor and the frequency reference in the same optic, and present the design of a self-referenced single-element dual-interferometer (SEDI) inertial sensor capable of reaching sub-picometer precision}.

In Section~\ref{section:compact-ifo} we introduce DFM, the key technology behind the SEDI experiment. Section~\ref{section:SEDI} presents the design of the SEDI inertial sensor. A structural analysis determining the experimental feasibility of the device is reported in Section~\ref{section:performance}. A noise analysis, including inherent imperfections, is then carried out and reported in Section~\ref{section:noise}. Finally, the results are summarized in Section~\ref{section:summary}.

\section{Deep Frequency Modulation Interferometry}
\label{section:compact-ifo}
There are several techniques to increase the working range of two beam interferometers~\cite{Watchi2018}. For example, heterodyne interferometry uses two laser frequencies to generate a beat signal from which the phase can be extracted via a phasemeter. Due to the frequency difference $\Omega$ between the two interfering fields in a heterodyne interferometer, the output power varies with time as
\begin{equation}
\label{eq:heterodyne-signal}
P_{\text{out}} \propto 1 + C \cos \left( \Omega t + \phi \right).
\end{equation}
where $\phi$ is the optical phase difference between the two arms, and $C$ is the interferometric contrast or fringe visibility. The amplitude and phase can be read out via an in-phase/quadrature (I/Q) demodulator phasemeter~\cite{Gerberding2015b} applied to the digitized photoreceiver output, which can be implemented, e.g., using a field-programmable gate array (FPGA). These two orthogonal signals can be used to extract the phase in post-processing.

However, the preparation of such laser beams with a frequency offset is not trivial and requires complex optics and electronics, and it is therefore not ideal for scaling to multi-DOF test mass (TM) read out. Meanwhile, optical fibers offer the most convenient way to deliver laser beams, but the length fluctuations of fibers can adversely affect phase measurements. \added{Several techniques can be used to cancel fiber-induced phase noise, such as using an offset phase-locked loop (e.g., as done in ~\cite{Chwalla2020}) or an active fiber phase noise cancellation system (e.g., as described in~\cite{Ma1994}), but again, these methods imply an added complexity of the optics and electronics setup}. DFM interferometry enables ultra-compact optical systems that are very robust to such fluctuations owing to the self-homodyning effect, whilst providing signals in a form similar to Equation~\ref{eq:heterodyne-signal} that can be read out via an I/Q demodulation phasemeter.

\begin{figure}[h]
\centering
\includegraphics{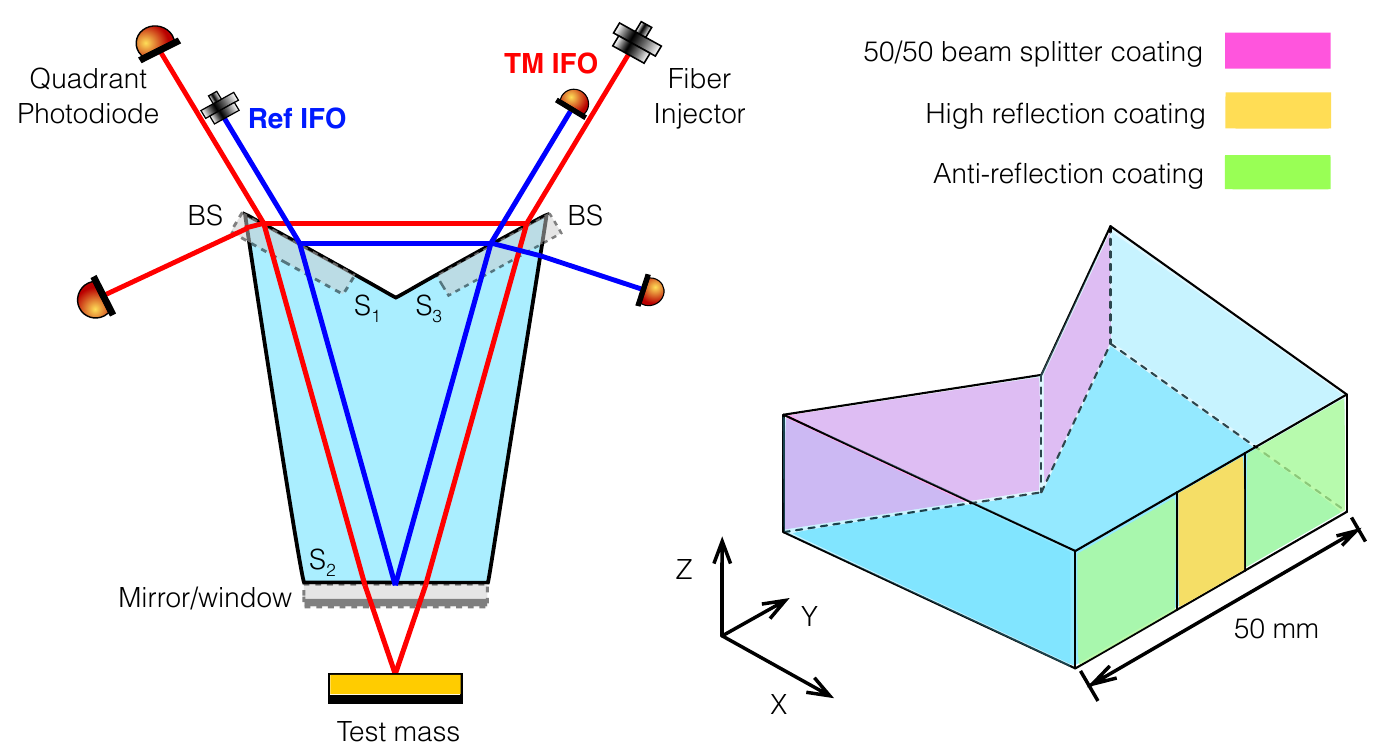}
\caption{Single-element dual-interferometer optical head layout. A frequency-modulated laser beam is split and delivered via two optical fibers to the heptagonal prism (a). The prism has three main optical surfaces: $S_1$ and $S_3$ are used to split and recombine the laser beams, while $S_2$ acts as mirror for the reference interferometer (Ref IFO), and as a window for the test mass inertial sensing interferometer (TM IFO). The shape of the prism is optimized via simulation to provide insensitivity to manufacturing tolerances and rejection of stray light noise. Also drawn is a 3D view of the prism showing the required optical coatings (b).}
\label{fig:crown_concept}
\end{figure}

By applying a strong sinusoidal frequency modulation to the laser input of an unequal arm-length interferometer, with modulation frequency $\omega_m = 2 \pi f_m$, modulation phase $\phi$, and modulation depth $\Delta f$, the output power takes the form~\cite{Gerberding2015}
\begin{equation}
\label{equation:DFM-signal}
P_{\text{out}} \propto 1 + C \cos\left(\phi + m \cos\left[ \omega_m t + \psi \right]  \right),
\end{equation}
where $m = 2\pi\Delta f \tau$ is the effective modulation depth, which depends linearly on $\Delta f$ and on \added{the light travel time shift $\tau = \Delta L / c$ between the two arms of the interferometer ($c$ is the speed of light and $\Delta L$ is the optical pathlength difference between the arms)}. The interferometer output is therefore periodic with $f_m$, and its waveform is dependent on the interferometric phase $\phi$. The voltage signal $v_{\text{out}}(t)$ from a photodiode is digitized after appropriate analog amplification and anti-alias filtering, and then demodulated with sine and cosine tones at $N$ harmonics of the modulation frequency before being low-pass filtered. The resulting demodulated signal for each harmonic is given by the quadrature $Q_n$ and in-phase $I_n$ components,
\begin{align}
Q_n &= v_{\text{out}}(t) \cos n \omega_m t \approx k  J_n(m) \cos \left( \phi + n \frac{\pi}{2} \right) \cos n \psi, \\
I_n &= v_{\text{out}}(t) \sin n \omega_m t \approx - k  J_n(m) \cos \left( \phi + n \frac{\pi}{2} \right) \sin n \psi.
\end{align}
\added{where $J_n(m)$ are the Bessel functions of the first kind.} Typically we use the first ten harmonics to fit the four parameters with a Levenberg-Marquardt (least-squares) routine or Kalman filter; $k$, $m$, $\psi$ and $\phi$.

\added{Apart from providing a high resolution and high dynamic range displacement measurement through the readout of the interferometric phase $\phi$, DFM interferometry can also provide a measurement of the absolute position of the TM through the effective modulation depth $m$ measurement, which encodes the delay in light travel time between the interferometer arms. This measurement can be performed, e.g., by making a separate measurement of the frequency modulation depth $\Delta f$. For example, for the case of a 500\,mm optical pathlength difference, and a typical relative readout accuracy of $m$ of $\Delta m / m \sim 4.5\cdot 10^{-6}$, the resulting absolute ranging accuracy is $\Delta(\Delta L) = \Delta L \cdot \Delta m / m \simeq 2.2\,\upmu$m. This measurement can be combined with the interferometric phase measurement to increase the working range of the detector, and possibly bridge the gap between the interferometer and ranging signal. In order to achieve the aforementioned fractional $m$ stability a control system applies a feedback signal to the laser controller acting on $\Delta f$, and thus stabilizing the effective modulation index~\cite{Isleif2018PhD}. }

\section{\added{Design of the SEDI Experiment}}
\label{section:SEDI}
DFM interferometry has been implemented previously on a single-element interferometer consisting of a custom-made triangular prism capable of sensing the motion of a test mass with sub-picometer precision~\cite{Isleif2019}. The small volume of this single-element interferometer, hereafter referred to as ``optical head'' (OH), means that it is readily scalable to interrogate one or multiple test masses in several degrees of freedom. While the triangular prism OH setup is very compact, with a volume of just a few cubic inches, it relies on a second, separate interferometer, also employing DFM readout, for laser frequency pre-stabilization~\cite{Gerberding2017}. 

For the next generation of these devices, we incorporate both the inertial sensor and the reference interferometer in the same optic. This is made possible by a custom-design prism (Figure~\ref{fig:crown_concept}). Using optical fibers we split and deliver a single frequency-modulated laser signal to one or several of these OH, enabling us to sense the motion of a system in multiple degrees of freedom, and eliminating the need for a separate frequency reference. Hence, a single-element dual-interferometer (SEDI) inertial sensor is realized.

\begin{figure}
\centering
\includegraphics{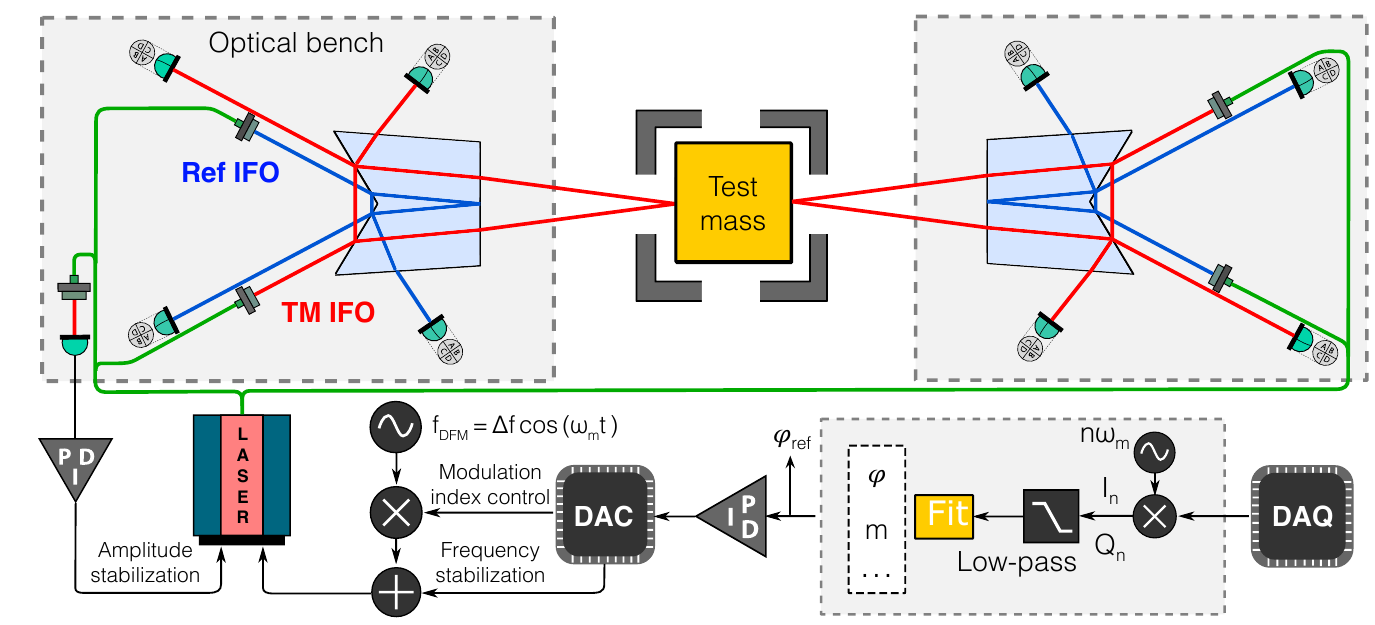}
\caption{\added{Design layout of the SEDI experiment}. A cubic test mass is probed from opposite sides by a pair of heptagonal prism optical heads that are fed from a single frequency-modulated laser source. Each optical head hosts two interferometers, one reading the displacement of the test mass and another acting as frequency reference. The DFM interferometric signals are captured by photodiodes, digitized, and processed by a phasemeter. A fit algorithm provides the differential phase measurement of each detector. The phase measurement of the reference interferometer is used as control signal for the laser's frequency and modulation index, and as calibration signal for the test mass displacement measurement.}\label{fig:SEDI_Experiment_Layout}
\end{figure}

There are many possible prism geometries capable of accommodating two unequal arm-length interferometers. We settle for a heptagonal prism with three main optical surfaces ($S_{1,2,3}$). This seamless piece of fused silica glass, obtained via optical contacting of smaller parts, hosts a test mass interferometer (TM IFO) and a reference interferometer (Ref IFO), using surfaces $S_1$ and $S_3$ for laser beam splitting and recombination. Surface $S_2$ features two different coatings and serves a double purpose: an inner portion of the surface is HR-coated to act as a mirror for the Ref IFO, and the remainder of the surface is AR-coated to serve as a transparent window for the TM IFO. The shape of the prism is completed by four lateral surfaces, and is optimized via numerical simulation to provide insensitivity to machining imperfections and maximal rejection of stray light noise.

The reference interferometer serves two purposes. First, it is used to compensate the very slow laser frequency drift by feeding back a control signal to the source. Secondly, it is used to remove the coupling of the remaining laser frequency noise from the TM IFO measurements, which appears as an unavoidable source of noise due to the unequal arm-length. The latter is done by combining the differential phase measurements of both interferometers in post-processing,
\begin{eqnarray}
\phi_{\rm tm} &=& 2\delta \phi+2\pi\delta f_{0}\tau_{\rm tm} + \sigma_{\rm tm} \label{eq:tm_sig}, \\
\phi_{\rm ref} &=& 2\pi\delta f_{0}\tau_{\rm ref} + \sigma_{\rm ref}, \label{eq:ref_sig}\\
\phi_{\rm tm} - \frac{\tau_{\rm tm}}{\tau_{\rm ref}}\phi_{\rm ref} &=& 2\delta \phi + \sigma_{\rm tm} + \frac{\tau_{\rm tm}}{\tau_{\rm ref}}\sigma_{\rm ref},  \label{eq:tm_sig_calib}
\end{eqnarray}
where $\delta \phi$ is the phase shift due to the perceived TM longitudinal motion, scaled by a factor of approximately $2$ due to the reflection setup; $\delta f_{0}$ is the laser frequency noise; $\tau_{\rm tm}$ and $\tau_{\rm ref}$ are the time delays due to the geometric optical pathlength difference between the short and long arms in the TM IFO and Ref IFO respectively (i.e., $\tau = \Delta L/c$, where $\Delta L$ is the optical pathlength difference between the arms, and $c$ is the speed of light); and the $\sigma_{\rm tm}$ and $\sigma_{\rm ref}$ terms represent additional noise sources. The cancellation of the frequency noise is limited by the accuracy of the $\Delta L$ measurements, obtained by measuring the effective modulation index $m$. \added{Measurements using the same optics, electronics, and readout algorithm~\cite{Isleif2018PhD} show a fractional effective modulation index stability of $\Delta m / m \approx 4.5 \cdot 10^{-6}$, leading to a typical error in the estimation of $\Delta L$ of $\sim 1\,\upmu$m for both interferometers.}

With frequency noise suppressed by combining the signals from the TM and reference interferometers, the main noise sources are thermal (i.e., thermal expansion of the optic, refractive index variations, and fiber injector jitter), electronic (i.e., phasemeter noise), and optical (i.e., stray light noise, and cross coupling of test mass tilt into the length measurement) in nature. All of these sources are investigated in Section~\ref{section:noise}, with the exception of stray light noise, that is tackled in Section~\ref{section:performance}. \added{The sensor is expected to be applied in vacuum, and to be well isolated from external forces.}

Figure~\ref{fig:SEDI_Experiment_Layout} shows an experimental layout where a test mass is being probed from opposite sides by a pair of SEDI detectors. The same layout still applies to any number of SEDI sensors probing any number of degrees of freedom of one or multiple test masses, simply by scaling the number of required fiber injectors and phasemeter channels. As we show in Section~\ref{section:noise}, this setup allows a drastic reduction of thermal noise, resulting in an ultimate performance limited by electronic noise and tilt-to-length coupling noise, two sources that can be considered fundamental for such measurement schemes.

\added{To reach sufficient modulation depth in centimeter-scale setups such as SEDI, the frequency modulation applied to the laser beam must be at the GHz level, which can make the search for a suitable laser source slightly challenging. We have opted for the commercially available 1064.5\,nm external cavity diode laser TLB 6821 from Newport, which we have verified experimentally to provide sufficient modulation strength for our purpose. The photodiodes are based in InGaAs with a small active area diameter of 0.5\,mm to achieve high bandwidth and interferometric contrast, and a trans-impedance amplifier is applied to each photocurrent signal channel. The signals are subsequently digitized using a data acquisition system with a sample rate of 250\,kHz, and the digital signals are demodulated by a software phasemeter that decimates the data rate down to 100\,Hz and implements the non-linear fit algorithm or Kalman filter.}


\section{\added{Structural Analysis of the Optical Head Geometry including Manufacturing Imperfections}} \label{section:performance}

\begin{figure}
\centering
\begin{tikzpicture}
\draw (0,0) node[inner sep=0] {
\includegraphics[trim={4.25cm 3cm 5.5cm 2.5cm},clip,width=7cm]{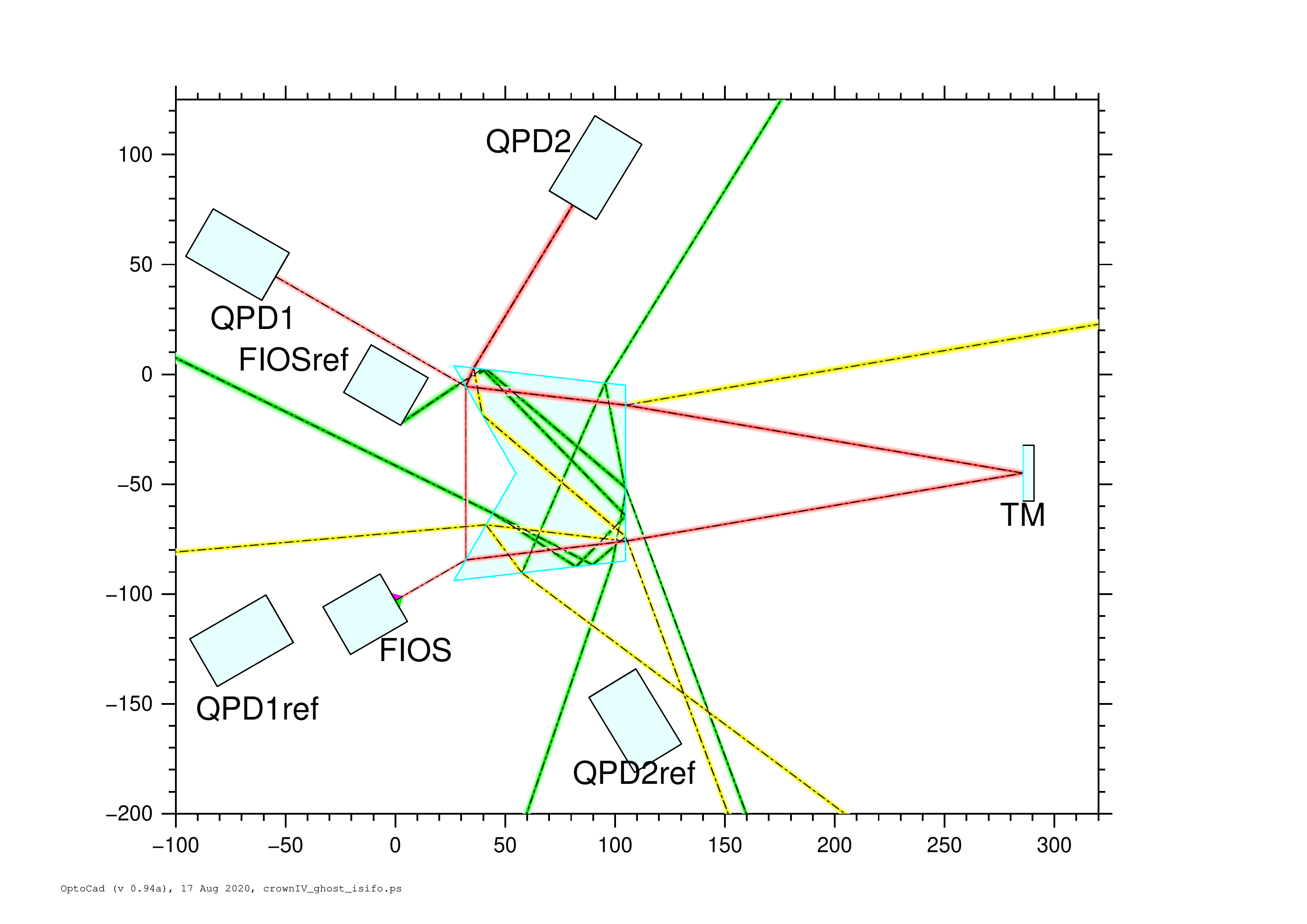}
};
\draw (-2,2) node {(a)};
\end{tikzpicture}
\begin{tikzpicture}
\draw (0,0) node[inner sep=0] {
\includegraphics[trim={4.25cm 3cm 5.5cm 2.5cm},clip,width=7cm]{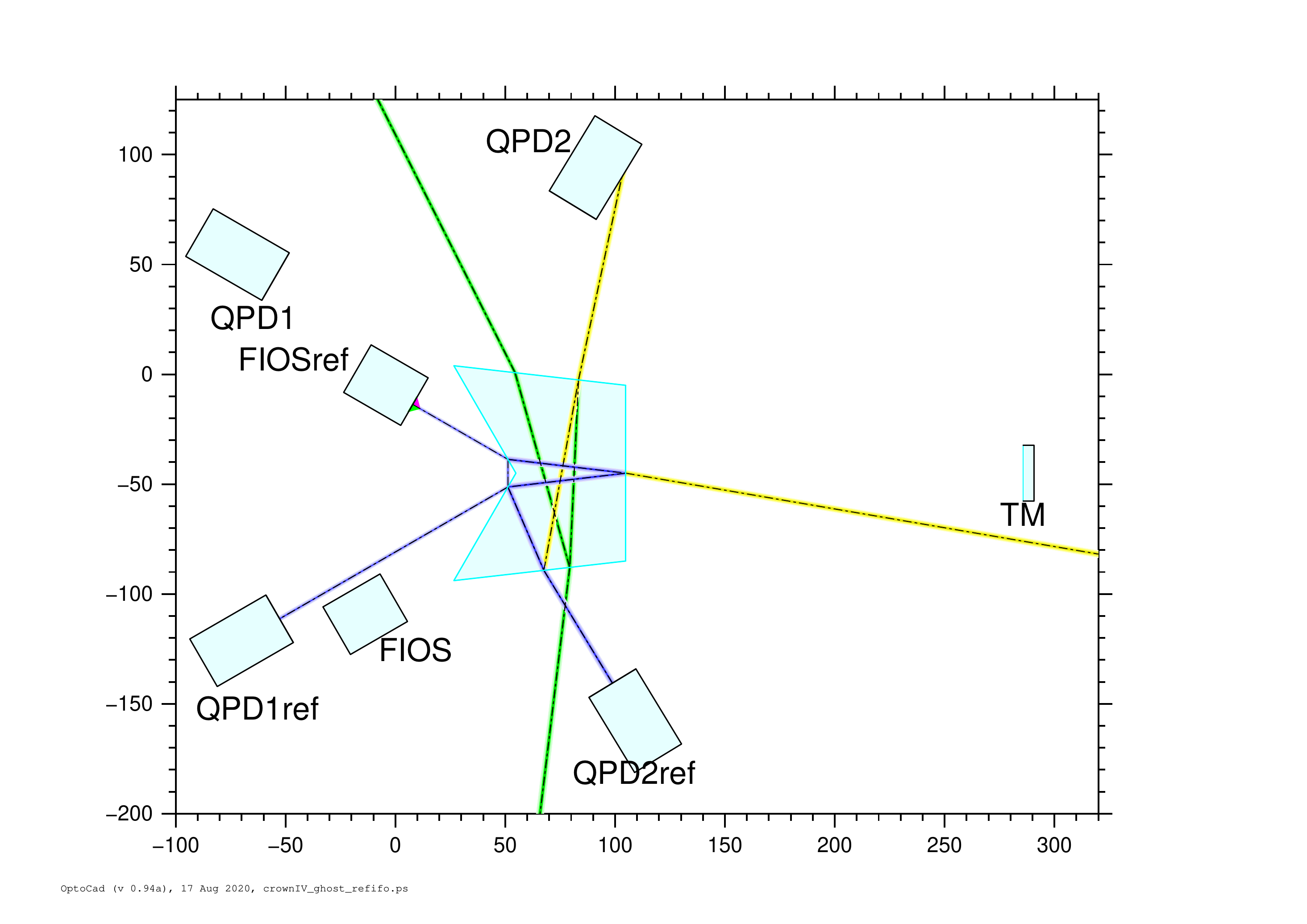}
};
\draw (-2,2) node {(b)};
\end{tikzpicture}
\caption{Ghost beam suppression in TM IFO (a) and Ref IFO (b). Beams resulting from unwanted reflections or transmissions in the optic may interfere with the nominal interferometric beams and spoil the measurement. The relative angle between the side surfaces is optimized via simulation to yield a configuration where ghost beams down to a certain power threshold are kept from impinging the detectors. Shown in red and blue are the nominal beams of the test mass and the reference interferometer, respectively, in the already optimized optical head. Absolute errors of $10^{-2}$ are introduced in the power transmission coefficient of each surface, and ghost beams down to a power threshold of $10^{-12}$ relative to the nominal beam power are considered. In this plot we represent ghost beams at the $10^{-3}$ (yellow) and $10^{-7}$ (green) relative power levels for the sake of clarity, see the appendix for a more complete picture.}
\label{fig:ghost}
\end{figure}

For reaching displacement sensitivities on the order of $1\,\text{pm}/\sqrt{\text{Hz}}$ or lower, it is essential to keep stray light noise in check, a lesson learnt from experiments for LISA performed on ground (e.g., see~\cite{Isleif2018}). Beams resulting from unwanted reflections or transmissions, known as ``ghost beams'', inevitably occur due to imperfections in the optical surfaces of the interferometer, and will spoil the measurement when captured by the detectors. A strategy to reduce this source of noise is imperative in order to reach a sub-picometer performance. By using a custom optic, we have the opportunity to produce a specially engineered shape where ghost beams are dealt with accordingly so as to keep them from affecting the measurement.

A disadvantage of any experiment using complex optical elements is dealing with manufacturing tolerances and imperfections. For example, relative alignment errors between the three optical surfaces of the OH can cause a bad overlap between the interferometer arms and lead to poor interferometric contrast. Similarly, the perpendicularity of the optical surfaces with respect to the base of the prism is essential to maintain both interferometers in-plane, as small deviations from such perpendicularity can cause the beams to veer off-plane significantly.

To simulate these effects and optimize the prism geometry, we build an optical model of the experiment using the interferometer simulation software Ifocad~\cite{Ifocad}. The OH geometry is parametrized and included in the model along with all the features depicted in Figure~\ref{fig:SEDI_Experiment_Layout}. Ifocad provides proven methods for propagating general astigmatic Gaussian beams in 3D space, as well as for computing most relevant interferometric signals, such as the interference contrast, the longitudinal pathlength signal (LPS), differential wavefront sensing (DWS) signals, and differential power sensing signals (DPS)~\cite{Wanner12OC}. For a single element photodiode the LPS is computed as
\begin{equation}
	s_{\text{LPS}} \equiv \frac{\phi}{k} = s + \frac{1}{k}\arg\left(\iint  E_1 E_2 ^{\ast} dS_{\text{pd}} \right), 
	\label{equation:LPS}
\end{equation}
where $\phi$ is the interferometric phase, $k=2\pi/\lambda$ is the wave number, $s$ is the macroscopic accumulated optical pathlength difference during beam propagation through the setup (which in this setup is in the order of $\sim 100\,$mm, see Table~\ref{tab:geo_info}), $E_1$ and $E_2$ are the complex amplitudes of the interfering beams projected into the detector plane, and $dS_{\text{pd}}$ is a surface element in the detector surface. There are several definitions of the LPS for a quadrant photodiode (QPD)~\cite{Wanner2015}, and in this paper we use the ``average phase'' definition, where the LPS is computed as $s_{\text{LPS}} = (\phi_A+\phi_B+\phi_C+\phi_D)/4k$, where $\phi_{A\dots D}$ are the interferometric phases measured by each of the QPD segments. The computation of $s_{\text{LPS}}$ therefore takes into account any effects stemming from the transverse distributions of the two interfering beams in the detector plane, as well as Gouy phase shifts, which have been identified as potential sources of error in precision interferometers~\cite{Yoshino2020}.

The nominal parameters used in the simulation are listed in the appendix. Some important geometrical parameters are given in Table~\ref{tab:geo_info}. For the results presented in this paper we compute the interferometric signals of the QPD on the reflection port of each interferometer (QPD1 and QPD1ref for the TM IFO and Ref IFO respectively).

\begin{table}
\centering
\small
\caption{\label{tab:geo_info} Geometrical parameters relevant to the structural analysis and noise investigations. AOI: angle of incidence.} 
\begin{tabular}{@{}lclclclcl}
\bhline{1.0pt}
Parameter & Value\\ \hline
TM IFO intra-prism pathlength (mm) & 212.53\\ 
REF IFO intra-prism pathlength (mm) & 156.55\\
TM IFO arm-length difference (mm) & 500.55\\
REF IFO arm-length difference (mm) & 143.98\\
$S_2$ to TM surface distance (mm) & 180.88\\
AOI to the TM (deg) & 9.71 \\\bhline{1.0pt}
\end{tabular}
\end{table}

An automatic beam tracing routine is used to propagate the beams within the optical system formed by the fiber injectors, the OH, and the QPDs (Figure~\ref{fig:ghost}). We introduce an absolute error on the power transmission coefficient of each coating of $10^{-2}$ as a worst case scenario, and consider all resulting beams down to a certain power threshold. By sweeping the parameter space of possible prism geometries, we find that the relative angle between the two side surfaces of the prism is by far the biggest driver of the amount of stray light directed towards the detectors. By tuning this angle and inspecting the resulting set of ghost beams it is possible to choose a geometry that guarantees suppression of stray light to a very large degree. 

A ghost beam having perfect overlap with the nominal beam and having a power above $3.5 \cdot 10^{-11}$ relative to the nominal beam power could cause instabilities at the picometer displacement level~\cite{Isleif2018PhD}. Hence, we aim to maintain ghost beams with relative powers down to $10^{-12}$ from impinging the detectors, therefore reducing stray light noise significantly below the total noise floor (see Figure~\ref{fig:ghost_low} in appendix). This optimization is performed in addition to other means of dealing with stray light, such as the placement of suitable apertures or beam dumps that can tackle a particular set of beams.

\begin{figure}
\centering
\begin{tikzpicture}
\draw (0,0) node[inner sep=0] {\includegraphics{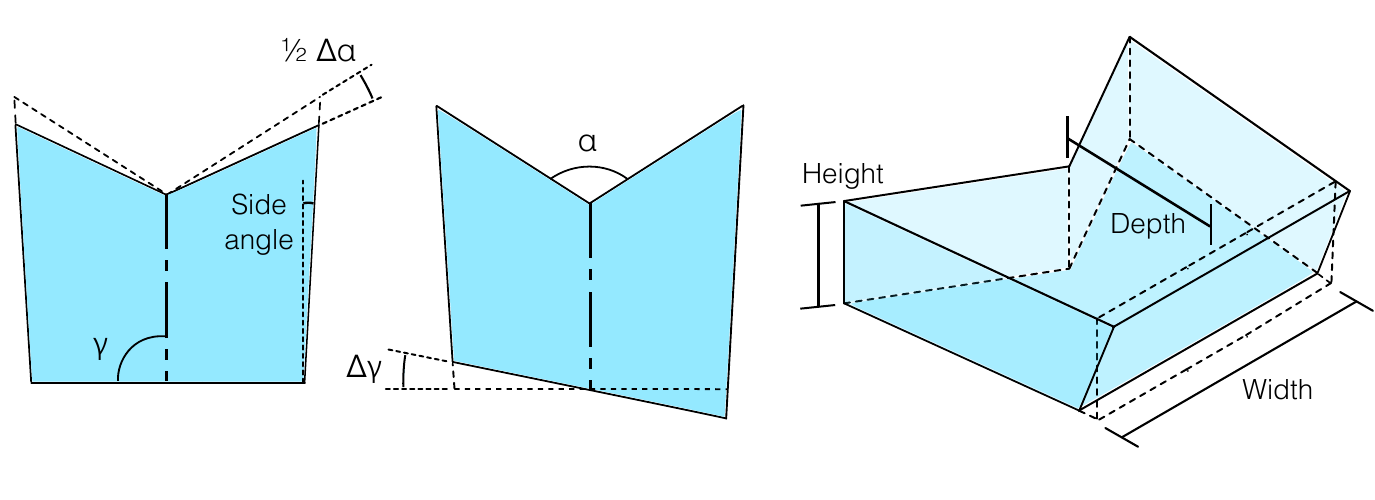} };
\draw (-6,2) node {(a)};
\draw (3,2) node {(b)};
\end{tikzpicture}
\caption{Critical manufacturing tolerances are divided into two categories. Tolerances affecting the relative alignment between optical surfaces (parametrized by $\alpha$ and $\gamma$) lead to in-plane beam misalignments that can be compensated for in both interferometers by tuning the direction of the incident beams; tolerances affecting the perpendicularity of the optical surfaces with respect to the prism base  cause off-plane beam misalignments and result in an unavoidable loss of interferometric contrast.}
\label{fig:manu_error}
\end{figure}

To analyze the sensitivity of the detector to manufacturing tolerances of the prism, we inject different types of geometrical errors into the OH model. Among the geometrical parameters that are critical to the performance of the device, we distinguish between two types: relative alignment errors between the optical surfaces, parametrized by angles $\alpha$ and $\gamma$ (Figure~\ref{fig:manu_error}a); and deviations from perpendicularity of the optical surfaces relative to the base of the prism (Figure~\ref{fig:manu_error}b). The former couples to beam misalignments on the $xy$ plane, and may be compensated to some degree by realigning the fiber injectors; the latter causes beams to veer off-plane and its compensation is unfeasible.

\begin{figure}[h]
\includegraphics{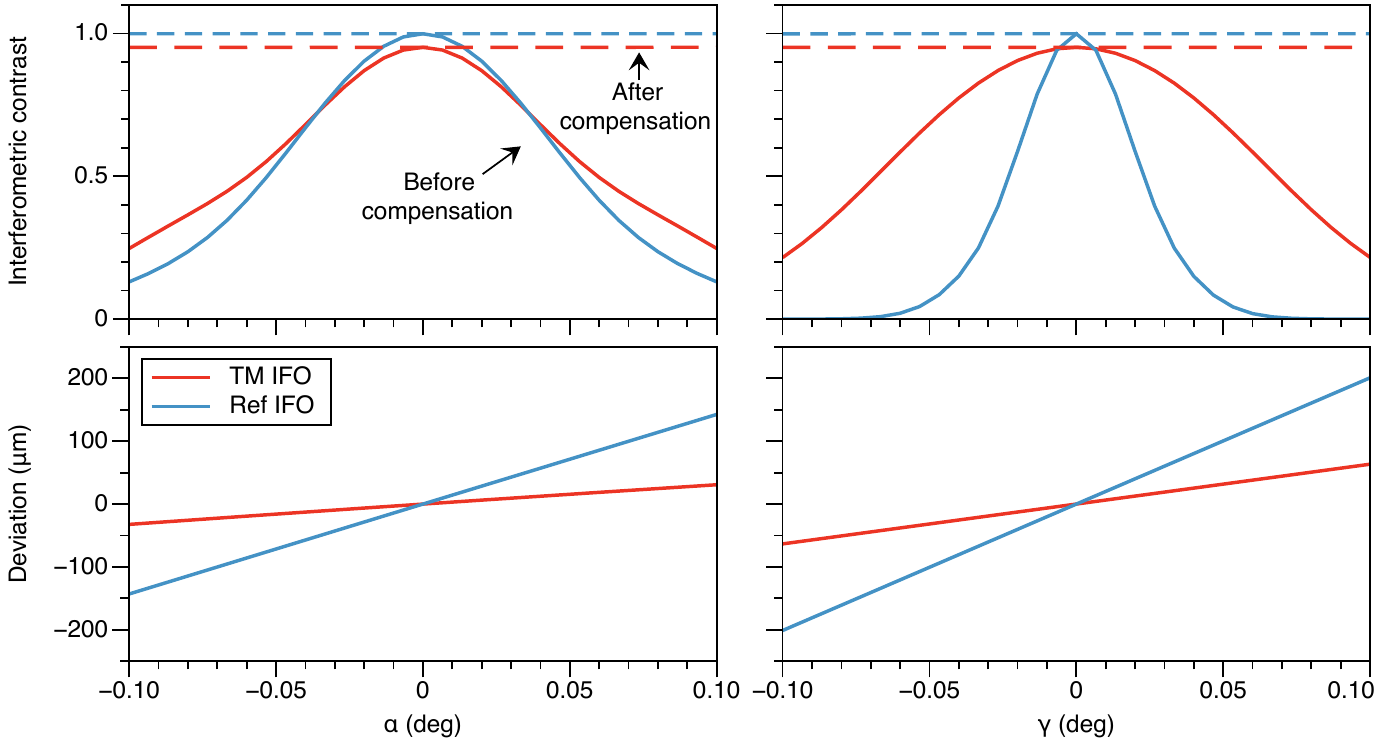}
\caption{Sensitivity to manufacturing imperfections affecting the relative alignment between optical surfaces, as parametrized by the $\alpha$ (left) and $\gamma$ (right) angles. The top plots show the interferometric contrast as a function of the deviation from the nominal geometry for both interferometers with and without compensation by altering the alignment of the input beam direction. The input beam direction is tuned by translating the fiber injector by up to 1\,mm, and rotating it by up to 0.5\,degree. The bottom plots show the deviation from the nominal optical pathlength difference between the arms in both interferometers due to the required compensation.}
\label{fig:h_tolerance}
\end{figure}

In the $xy$-plane of the prism, angle $\alpha$ sets the relative angle between the beam splitting and recombination surfaces $S_1$ and $S_3$, while angle $\gamma$ sets the relative angle between the normal of the mirror/window surface $S_2$ and the nominal symmetry axis of this cut plane (Figure~\ref{fig:manu_error}). We introduce deviations of these angles in the model and compute the resulting degradation of the interferometric contrast. We find that it is possible to completely compensate for these types of errors in both interferometers by fine-tuning the direction of the incident beams (Figure~\ref{fig:h_tolerance}). Hence, the detector is insensitive to manufacturing tolerances affecting the relative alignment between optical surfaces.

On the other hand, manufacturing imperfections affecting the perpendicularity of the optical surfaces with respect to the base of the prism are more difficult to compensate for. The input beam couplers are able to introduce small vertical beam angles, as well as small vertical beam shifts. Typically, however, this is avoided at all costs, since it requires rotating the lens held in the fiber coupler assembly, and therefore introduces astigmatism; it also tends to introduce strong couplings between some of the DOFs of the fiber coupler which are very difficult to control during construction. The in-plane interferometer design therefore takes priority, and we set a tolerance specification on the perpendicularity of the optical surfaces based on the degradation of the interferometric contrast that we deem allowable. 

\begin{figure}[h]
\centering
\includegraphics{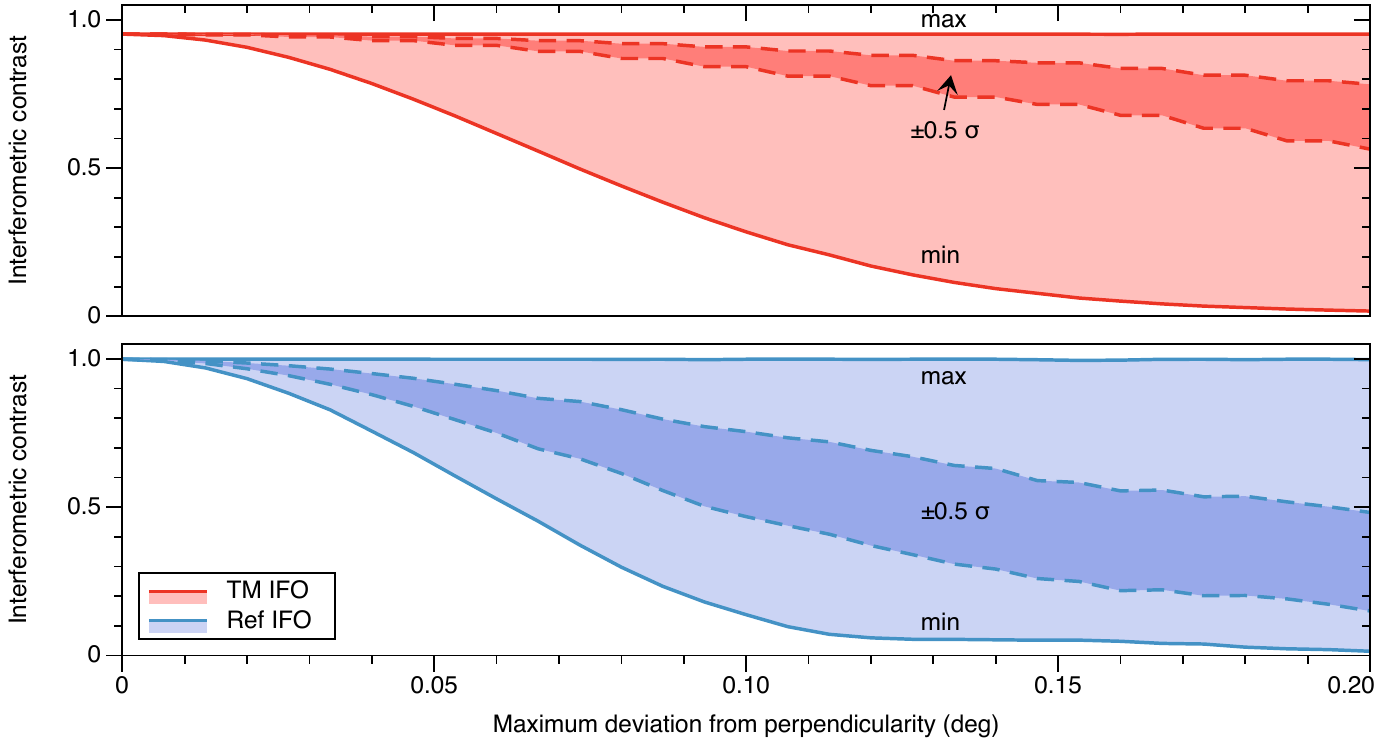}
\caption{Sensitivity to manufacturing imperfections affecting the perpendicularity of the optical surfaces. The plots show the degradation of the interferometric contrast as a function of the amplitude of the uniform distribution of deviations from perpendicularity injected into all three optical surfaces of both interferometers. The number of Monte Carlo samples at each point is 1000. The dark-shaded area bordered by dashed lines corresponds to $\pm 0.5\,\sigma$, while the light-shaded region is bordered by continuous lines representing the maximum and minimum of the distribution.}
\label{fig:v_tolerance}
\end{figure}

To properly characterize this effect, we perform a Monte Carlo simulation where we inject errors in the form of deviations from perpendicularity into all three optical surfaces following a uniform distribution, and compute the resulting interferometric signals (Figure~\ref{fig:v_tolerance}). We find that TM IFO performs better than Ref IFO in this regard, as expected, due to the fact that $S_2$ acts as a window for the TM IFO and its perpendicularity has little effect on this interferometer. Simulations show that deviations of up to $0.1$ degrees from perpendicularity are allowable whilst maintaining the interference contrast greater than 14\,\% in Ref IFO, and greater than 28\,\% in TM IFO. We verified that the DFM fitting algorithm still performs to the required level using signals with this level of contrast. Based on discussions with the precision glass machining companies we contacted, this perpendicularity requirement is feasible.

\section{\added{Noise Analysis for Future Experiments using 1 or 2 Optical Heads}} \label{section:noise}

Having determined the feasibility of manufacture, we turn our attention to the main noise sources of the SEDI inertial sensor. The largest entry in the noise budget, that emerges from having an interferometer of unequal arm lengths, is the laser frequency noise. This source of noise is mitigated by the dual interferometer configuration (see Equation~\ref{eq:tm_sig_calib}). Figure~\ref{fig:SEDI_Experiment_Layout} shows the control loop for the laser frequency. Assuming the laser frequency is only stabilized below the observation band to maintain the laser operating point, the free-running noise is used in the rest of this section.

The second entry in the noise budget is thermal noise, which can be very severe in optical setups employing large continuous optics like the SEDI sensor. Thermal noise is generally compound of two effects: mechanical deformation due to thermal expansion and refractive index variations. The latter is specially relevant in this experiment due to the interferometers' long arms having significant intra-prism optical pathlengh, as listed in Table~\ref{tab:geo_info}. Since thermal noise and laser frequency noise are uncorrelated, it is generally not possible to cancel both simultaneously.

We also calculate the displacement noise stemming from the angular and translational jitter noise of the fiber injector optical subassemblies (FIOS), which makes the next entry in the noise budget. Lastly, we also obtain the coupling between the angular motion of the test mass and the displacement measured by the sensor, or tilt-to-length (TTL) coupling, which takes the value of $0.06\,\text{pm}/\upmu\text{rad}$ assuming a perfectly manufactured prism.

\begin{table}[]
\small
\centering
\caption{\label{tab:freq_thermal_noise} Amplitude of the displacement noise of TM IFO and Ref IFO in a single optical head due to the main noise sources. The coefficient of thermal expansion and the $dn/dT$ term of fused fused silica are taken to be $0.55 \cdot 10^{-6}$\,K$^{-1}$, and   $9.6 \cdot 10^{-6}$\,K$^{-1}$, respectively.} 
\begin{tabular}{lccc}
\toprule
noise source    			 & magnitude 		&  TM IFO  &  Ref IFO \\
\midrule
frequency noise 			 & 100\,MHz       		& $1.78\times10^{5}$\,pm & $5.11\times10^{4}$\,pm\\ 
refractive index fluctuation & $1\times10^{-5}$\,K & $1.41\times10$\,pm & $1.04\times10$\,pm\\
thermal deformation			 &$1\times10^{-5}$\,K  & $5.21\times10^{-1}$\,pm & $7.92\times10^{-1}$\,pm \\
FIOS jitter: pitch 			 & 1.0\,urad  &$4.00\times10^{-2}$\,pm & $6.76\times10^{-3}$\,pm \\
FIOS jitter: yaw 		     & 1.0\,urad & $1.57\times10^{-1}$\,pm & $3.30\times10^{-2}$\,pm \\
FIOS jitter: displacement    & 10.0\,nm & $1.14\times10^{-4}$\,pm & $5.68\times10^{-5}$\,pm \\
TM tilt-to-length coupling   & 20\,nrad  & $1.14\times10^{-3}$\,pm & - \\
phasemeter noise~\cite{Isleif2018PhD}  & 0.4\,urad & $6.77\times10^{-2}$\,pm & $6.77\times10^{-2}$\,pm\\
\bottomrule
\end{tabular}
\end{table}

Table~\ref{tab:freq_thermal_noise} lists the resulting displacement noise amplitudes due to the different noise sources affecting the phase measurement in each interferometer, assuming the typical noise floor values also listed in the table. \added{The temperature noise floor is in line with the requirements achieved at the optical components level in space-based experiments such as LISA Pathfinder~\cite{Gibert2015}, and it is also in line with the performance that can be achieved in experiments performed on the ground in medium-to-high vacuum~\cite{Gerberding2017}.} 

As the experiment is scaled, however, we may take advantage of signal combinations with a lower total noise floor. For example, by probing a test mass from opposite sides as depicted in Figure~\ref{fig:SEDI_Experiment_Layout}, we obtain four interferometric signals:
\begin{eqnarray}
\phi^{L}_{\rm tm} &=& 2\delta \phi+2\pi\delta f_{0}\tau^{L}_{\rm tm} + \frac{2\pi}{\lambda_{0}}\Theta^L_{\rm tm}\delta T^L + \zeta_{\rm tm} \label{eq:phi_l_tm},\\
\phi^{R}_{\rm tm} &=& -2\delta \phi+2\pi\delta f_{0}\tau^{R}_{\rm tm} + \frac{2\pi}{\lambda_{0}}\Theta^R_{\rm tm}\delta T^R + \zeta_{\rm tm}+2\epsilon_{\rm tm} \label{eq:phi_r_tm},\\
\phi^{L}_{\rm ref} &=& 2\pi\delta f_{0}\tau^{L}_{\rm ref} + \frac{2\pi}{\lambda_{0}}\Theta^L_{\rm ref}\delta T^L + \zeta_{\rm ref} \label{eq:phi_l_ref},\\
\phi^{R}_{\rm ref} &=& 2\pi\delta f_{0}\tau^{R}_{\rm ref} + \frac{2\pi}{\lambda_{0}}\Theta^R_{\rm ref}\delta T^R + \zeta_{\rm ref}+2\epsilon_{\rm ref} \label{eq:phi_r_ref},
\end{eqnarray}
where
\begin{equation}
\Theta^i_j = \Theta_j + \Delta \Theta^i_j = a^i_j(\Delta l) + b^i_j(\Delta n)  \label{eq:thermo_coupling}
\end{equation}
is the coupling coefficient of temperature fluctuations $\delta T$ to the pathlength signal of the interferometer labeled by $j=\{\text{tm,ref}\}$ and $i=\{\text{L,R}\}$. This coupling coefficient consists of the two contributions $a^i_j(\Delta l)$ and $b^i_j(\Delta n)$, respectively, due to the thermal expansion and refractive index change of the optic. $\Theta_j$ is the nominal coefficient determined by design and $\Delta \Theta^i_j$ is a correction term, which is attributable, e.g., to compensations of the manufacturing imperfections discussed in the previous section. The $\zeta_j$ terms represent common-mode noises between the left (L) and right (R) interferometers, and the $\epsilon_j$ terms represent uncorrelated noise sources. The test mass motion signal $\delta \phi$ can be extracted by combining the two TM IFO signals,
\begin{eqnarray}
\phi^{A}_{\rm TM} &=& \left. (\phi^{L}_{\rm tm}-C_{\rm freq, tm}\phi^{R}_{\rm tm}) \middle/ 2(1+C_{\rm freq, tm}) \right. \nonumber \\
&\simeq&\delta \phi+\frac{\pi\delta f_{0}}{2}\left(\tau^{L}_{\rm tm}-C_{\rm freq, tm}\tau^{R}_{\rm tm}\right)+\delta \theta^A_{\rm tm}(\delta T) + \frac{\epsilon_{\rm tm}}{2}, \label{eq:phi_tm_a}
\end{eqnarray}
where 
\begin{equation}
C_{\rm freq, tm} = \frac{\tau^{L}_{\rm tm}}{\tau^{R}_{\rm tm}} = 1 + \Delta \bar{C}_{\rm freq, tm} + \Delta \hat{C}_{\rm freq, tm} \label{eq:cfreq}
\end{equation}
is the calibration factor for laser frequency noise suppression in this signal combination, and $\delta \theta^A_{\rm tm} (\delta T)$ is the residual thermal noise coupling. The calibration factor $C_{\rm freq, tm}$ is nominally unity assuming that the design and construction of the left and right OHs are identical; the term $\Delta \bar{C}_{\rm freq, tm}$ accounts for the deviation due to the compensation of manufacturing imperfections, estimated to be $O(10^{-3})$ from Figure~\ref{fig:h_tolerance}; the term $\Delta \hat{C}_{\rm freq, tm}$ accounts for the calibration accuracy of the arm-length difference, which is normally $O(10^{-5})$. Ideally, the residual frequency noise term is negligible without any calibration factor. However, the deviations of arm-length difference from the nominal values can potentially reach the order of 100\,$\upmu$m, which is not negligible. Therefore, $C_{\rm freq, tm}$ is introduced to reduce this noise source to the extent limited by the measurement accuracy of 1\,$\upmu$m of arm-length difference.

On the other hand, the similarity between the temperature fluctuations in the left and right OHs is in principle unknown, making thermal noise a potentially limiting noise source. Expanding the thermal noise coupling term from Equation~\ref{eq:phi_tm_a} yields
\begin{equation}
\delta \theta^A_{\rm tm}(\delta T) = \frac{\pi }{2\lambda_{0}}\left[\Theta_{\rm tm}\left(\delta T^L-C_{\rm freq, tm}\delta T^R\right) + \Delta\Theta^L_{\rm tm}\delta T^L - C_{\rm freq, tm}\Delta\Theta^R_{\rm tm}\delta T^R\right]. \label{eq:dels_delT}
\end{equation}
We now invoke the reference interferometer signals, expanding on the potential of the Ref IFO beyond the basic idea expressed in Equation~\ref{eq:tm_sig_calib}. We combine these signals with Equation~\ref{eq:phi_tm_a} in a way that allows the suppression of $\delta \theta^A_{\rm tm}(\delta T)$,
\begin{eqnarray}
\phi^{B}_{\rm TM} &=& \phi^{A}_{\rm TM} - \left. C_{\rm temp}\left(\phi^{L}_{\rm ref}-C_{\rm freq, ref}\phi^{R}_{\rm ref}\right) \middle/ 2(1+C_{\rm freq, tm}) \right. \nonumber\\
&\simeq& \delta \phi + \delta \theta^B_{\rm tm}(\delta T)+ \frac{\epsilon_{\rm tm}}{2} + \frac{C_{\rm temp}\epsilon_{\rm ref}}{2},\label{eq:phi_tm_b}
\end{eqnarray}
\added{where
\begin{equation}
C_{\rm temp} = \frac{\Theta_{\rm tm} }{\Theta_{\rm ref} } = \frac{ a_{\rm tm}(\Delta l) + b_{\rm tm}(\Delta n) }{ a_{\rm ref}(\Delta l) + b_{\rm ref}(\Delta n) }
\end{equation}
is a calibration factor determined by design}, and
\begin{eqnarray}
\frac{2\lambda_{0}}{\pi}\delta \theta^B_{\rm tm}(\delta T) &\simeq& \left(\Theta^L_{\rm tm}-C_{\rm temp}\Theta^L_{\rm ref}\right)\delta T^L - \left(C_{\rm freq,tm}\Theta^R_{\rm tm}-C_{\rm temp}C_{\rm freq,ref}\Theta^R_{\rm ref}\right)\delta T^R \nonumber\\
&\simeq& \left[\Delta\Theta^L_{\rm tm} - C_{\rm temp}\Delta\Theta^L_{\rm ref}\right]\delta T^L \nonumber\\
&\hspace{3mm}&- \left[\Delta\Theta^R_{\rm tm} + \Delta\bar{C}_{\rm freq,tm}\Theta_{\rm tm} - C_{\rm temp}\left(\Delta\Theta^R_{\rm ref} + \Delta\bar{C}_{\rm freq,ref}\Theta_{\rm ref}\right)\right]\delta T^R, 
\end{eqnarray}
is the residual thermal noise coupling where only the cross terms of $\delta T$ with $\Delta \Theta$ and $\Delta\bar{C}_{\rm freq}\Theta$ remain. \added{The term $C_{\rm temp}$ relates the thermally induced optical pathlength variations in the TM IFO to those in the Ref IFO, and is determined purely geometrically. Of course, the prisms are always imperfectly manufactured and assembled, and these imperfections are accounted for by the $\Delta \Theta$ terms.} In this way, the major thermal coupling which scales only with $\Theta$ can be eliminated from the signal regardless of the correlation between the temperature fluctuations in the two OHs. It should be noted that this method is valid because both TM IFO and Ref IFO are located in the same plane in the optic, which allows us to treat them as being subjected to the same temperature fluctuations. \added{This has been verified, as described henceforth}.

To obtain the spectrum of thermal displacement noise we build a finite element model of the OH. This allows us to analyze the response of the prism to temperature fluctuations of the environment, and to account for the thermal low-pass filter effect provided by the optic. A model of the 3D geometry of the prism is developed in COMSOL Multiphysics, where the equations of heat transfer in solids are solved in the frequency domain. The thermal coupling between the prism and its environment depends on the mechanism by which the prism is mounted (e.g., by optical contact to a glass baseplate or similar). Since the details of this mechanism are still unknown, we consider a worst-case scenario in which all of the prism's base surface area is perfectly coupled to the surrounding environment. A spectrum of random temperature fluctuations, relaxed towards lower frequencies, is applied uniformly to the base of the prism, and the resulting temperature response is measured throughout the prism.

The transfer function of temperature fluctuations is computed as a volume average for the whole prism, as well as locally along the interferometer arms. The former is used to compute the spectrum of displacement noise due to the thermal expansion of the prism, and the latter is used to compute the spectrum of displacement noise due to the local refractive index change. \added{We find that under these assumptions the temperature fluctuations in both interferometers are fully correlated within the frequency band of interest. In the interest of generality, however, we also analyze the situation where the correlation in temperature noise between TM IFO and Ref IFO follows an arbitrary shape, as we report in Figure~\ref{fig:combiB_cohTMRef}}.

The final performance of the SEDI inertial sensor (i.e., the residual noise in the TM IFO channel), is derived for a single OH and the dual-OH setup of Figure~\ref{fig:SEDI_Experiment_Layout}. To estimate noise floors we consider a realistic case where the optical heads have been imperfectly manufactured, and the consequent interferometer misalignment has been compensated by altering the alignment of the input beam couplers (i.e., a compensation such as those described in Figure~\ref{fig:h_tolerance}). These imperfections are injected in the model in such a way that the asymmetry between the interferometers becomes maximum, leading to minimum common-mode noise cancellation. 

We found that, due to these imperfections, the test mass TTL coupling could increase to the picometer level, and imperfections causing deviations from the nominal geometry of the kind that can be parametrized by $\gamma$ (see Figure~\ref{fig:manu_error}) were specially relevant to this effect. This deterioration is caused by a lateral shift of the point of incidence of the beam in the test mass arising from imperfections of this kind, which can be $40\,\upmu$m or more, and the consequent extra longitudinal pathlength added to the long arm of the TM IFO. This offset can be compensated by altering the alignment of the TM IFO at the expense of a minor reduction from the maximum contrast, or by laterally shifting the position of the test mass. We opt for the latter in this simulation, and assume that the correct positioning is not achieved, such that the test mass is laterally offset by $10\,\upmu$m from the nominal position.

\begin{figure}[htb]
\centering
\includegraphics{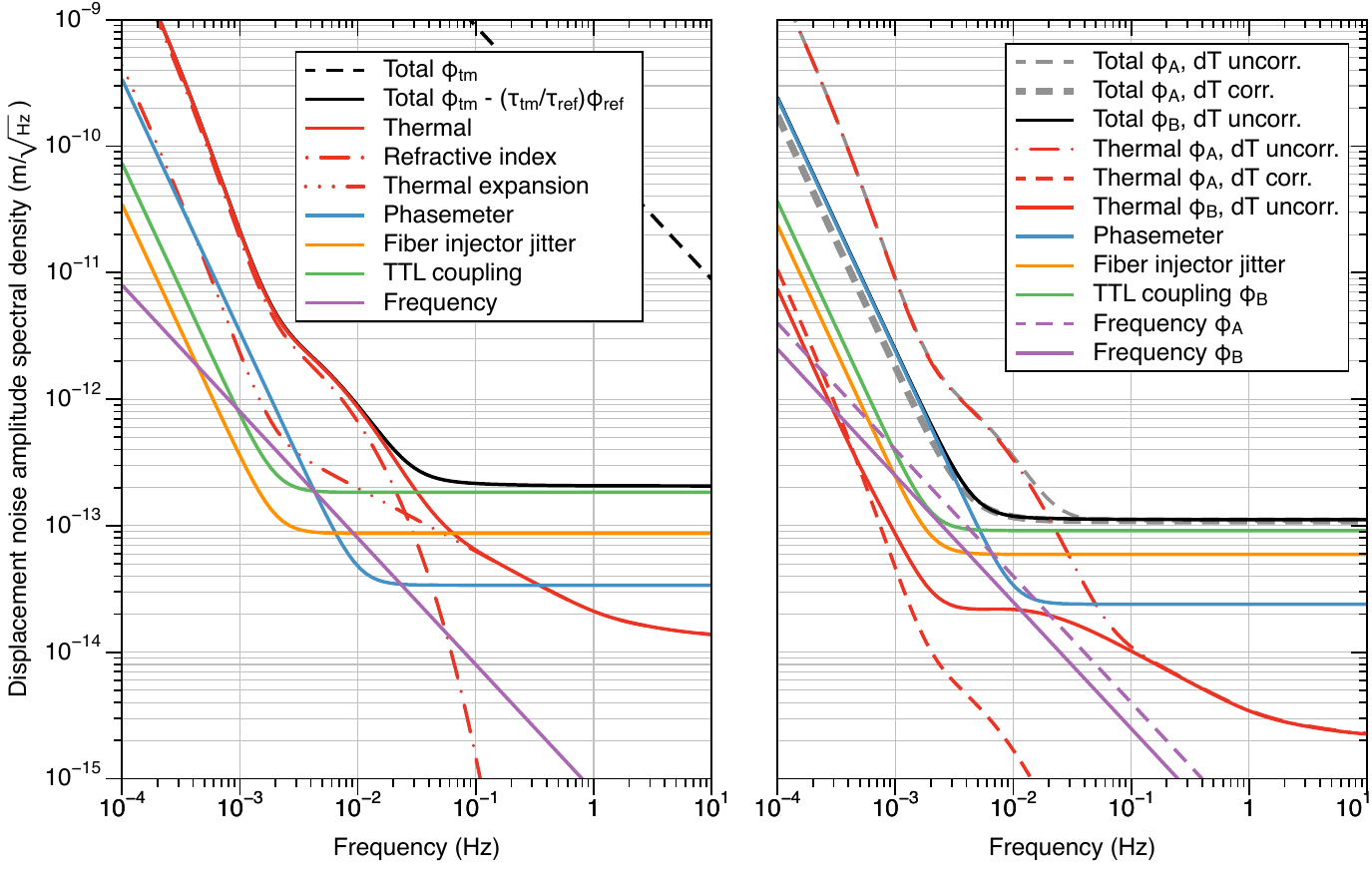}
\caption{\added{Simulated performance of the SEDI inertial sensor} using a single optical head (left), and two optical heads probing one test mass from opposite sides (right), in which the phase signals are written as a function of the correlation between temperature fluctuations in the two setups. These figures include characteristic imperfections of the manufactured prisms, and an absolute ranging error of 1\,um affecting the calibration factor. In the setup with two optical heads, two different calibrations are shown: $\phi_A$ and $\phi_B$ (see Equations~\ref{eq:phi_tm_a} and~\ref{eq:phi_tm_b} respectively), for distinct cases in which the interferometer temperature fluctuations are correlated or uncorrelated between the two optical heads used ($dT$~corr.\ and $dT$~uncorr.\ respectively).}
\label{fig:Residual_displacement_noise}
\end{figure}

The final results, including inherent imperfections in the construction, are shown in Figure~\ref{fig:Residual_displacement_noise}. The performance of a single OH is limited at low frequency by the residual thermal noise, while using two OHs allows for a phasemeter-noise-limited performance at low frequency, regardless of the coherence of the temperature fluctuation between the two devices. In both cases the sensor is limited at higher frequencies by TTL coupling noise. By applying the 
\begin{equation}
	10\,\frac{\text{pm}}{\sqrt{\text{Hz}}} \cdot \sqrt{1+\left(\frac{2\,{\rm mHz}}{f} \right)^{4}}
\end{equation}
displacement noise requirement of the LISA mission as the test mass displacement amplitude for frequencies $(20\,{\rm \upmu Hz}\leq f \leq 1\,{\rm Hz})$, 24\,dB signal-noise ratio (SNR) is achieved throughout the measurement frequency band. For frequencies above 100\,mHz, the SNR increases up to 40\,dB due to the decrease in thermal noise.

\section{Summary and Outlook}
\label{section:summary}

\added{We presented the design of the SEDI inertial sensor that is being built at the Albert-Einstein-Institute} to measure test mass displacements with a precision greater than $10^{-12}\,\text{m}/\sqrt{\text{Hz}}$ above 10\,mHz. The sensor fits in a small package of a few cubic inches, and offers this performance without the need for laser frequency pre-stabilization. The device consists of a single optical component that hosts two interferometers with arms of unequal length. The dual-interferometer configuration allows the SEDI sensor to act as its own reference for laser frequency noise suppression. Deep frequency modulation interferometry is applied as the technique of choice in this experiment as it enables the extreme simplicity of this setup, which is fed from a single frequency-modulated laser source.

Due to the intrinsic minimalism of this optical setup, this device can be integrated into an optical readout platform that features multiple optical heads and is able to interrogate several degrees of freedom of a mechanical system. We demonstrate how two of these optical heads can work together to suppress residual thermal noise and achieve sub-picometer precision above 2\,mHz, and up to the $100\,\text{fm}/\sqrt{\text{Hz}}$ level above 10\,mHz. The design of the optical head as well as its size pose a series of technological and scientific challenges, which are addressed in this article together with the devised solutions by means of optical simulations. A thorough noise analysis of the experiment is carried out, including manufacturing imperfections and their associated compensation mechanisms. The noise budget combines both analytical and computer models of the experiment. Furthermore, a finite element model of the optical head is used to compute the spectrum of temperature fluctuations in the interferometer.

\added{The perpendicularity of the three optical surfaces has been identified as the most critical manufacturing specification. This parameter has a large impact on the resulting interferometric contrast. This is a common issue in interferometers employing a single optical element. It was found that using smaller beams helps increase the likelihood of obtaining a good contrast from the manufactured prism. Another potential limitation of the device is that assembly requires accurate relative positioning of the fiber injectors and optical head in order to take full advantage of the stray light noise suppression strategy, and adjusting each fiber injector is a tedious job requiring a specialist with ample experience. An alternative to using two fiber injectors is using a fiber injector and a beamsplitter and mirror, which could simplify the assembly process at the expense of having to deal with additional ghost beams from the added beamsplitter. }

The potential applications of this new sensor cover many areas of science and technology. In experimental gravitational physics, and notably in next-generation space-based gravity missions, the SEDI sensor offers a scalable solution for multi-channel test mass readout. Satellites employing drag-free control to trail a free-floating test mass in order to follow an undisturbed geodesic require measuring the motion of said test mass in several degrees of freedom with high precision and large dynamic range. The reduced size and weight of the SEDI sensor means that several optical heads can be applied to offer increased sensitivity and redundancy in these measurements, making it a suitable candidate for these applications. 

The same measurement principle applies to inertial isolation systems on the ground, where the inertial stability of a reference mass can be transferred to an actively controlled platform through a high-gain feedback system. This is a key resource in many experiments that require suppression of seismic noise, notably in gravitational-wave detection, where the seismic activity at low frequency presents an important impediment to the discovery potential of ground-based detectors.

\added{Beyond space-based gravity missions or ground-based gravitational-wave detection experiments, the SEDI inertial sensor can find applications in fields across multiple disciplines. For instance, the first planned application of the sensor is its integration into the torsion balance facility being developed at the same institute. The inertial member of the torsion balance is suspended from a tungsten fiber and consists of a symmetric crossbar with cubic test masses at the end of each arm. Two SEDI sensors will be applied to opposite sides of one of the cubic test masses in a similar fashion to the case that is reported in this article (Figure~\ref{fig:SEDI_Experiment_Layout}). This will provide an out-of-loop measurement of the residual test mass motion that allows assessing the achieved performance.}

\section*{Appendix}
\begin{figure}
\centering
\begin{tikzpicture}
\draw (0,0) node[inner sep=0] {
\includegraphics[trim={4.25cm 3cm 5.5cm 2.5cm},clip,width=7cm]{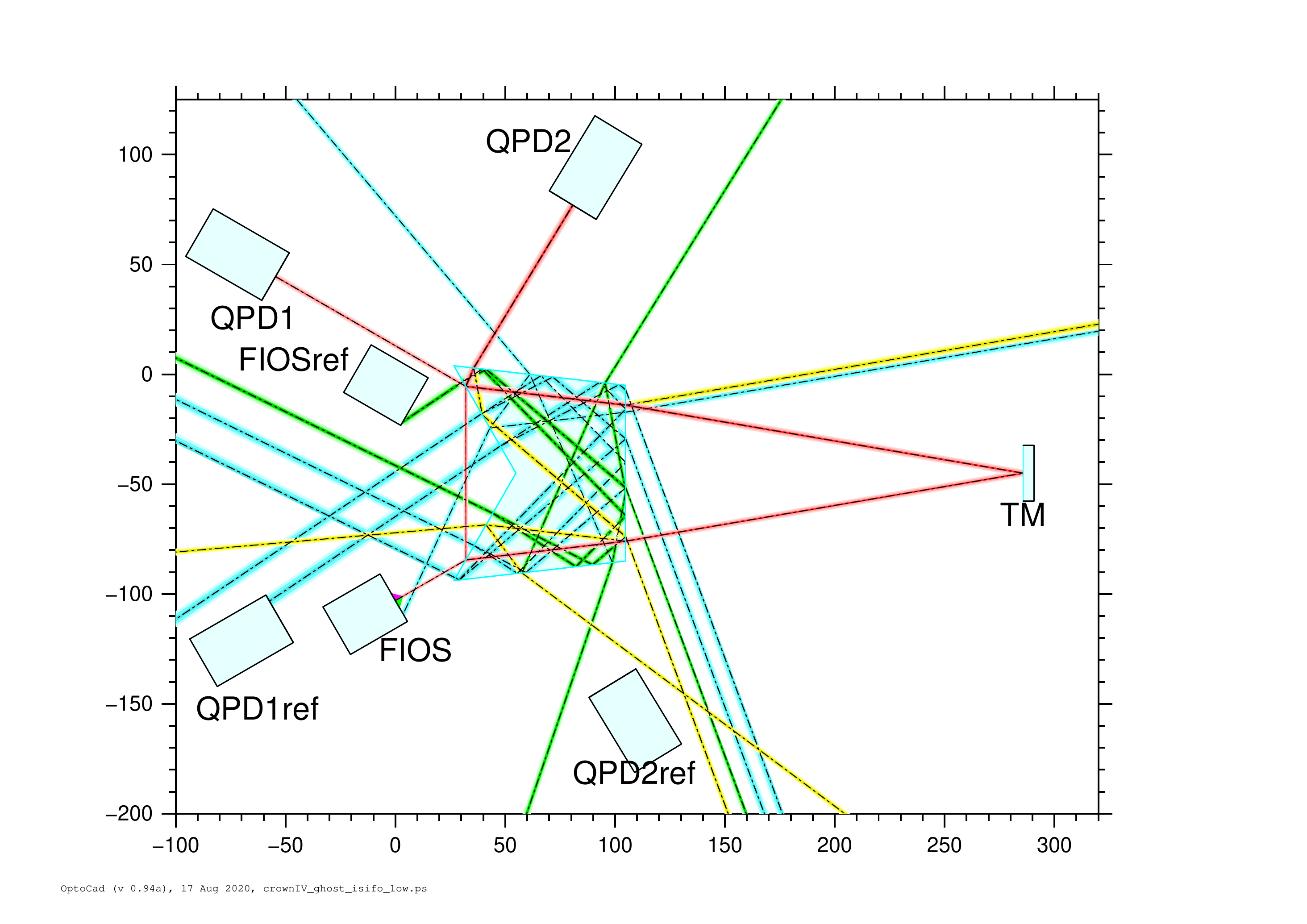}
};
\draw (-2,2) node {(a)};
\end{tikzpicture}
\begin{tikzpicture}
\draw (0,0) node[inner sep=0] {
\includegraphics[trim={4.25cm 3cm 5.5cm 2.5cm},clip,width=7cm]{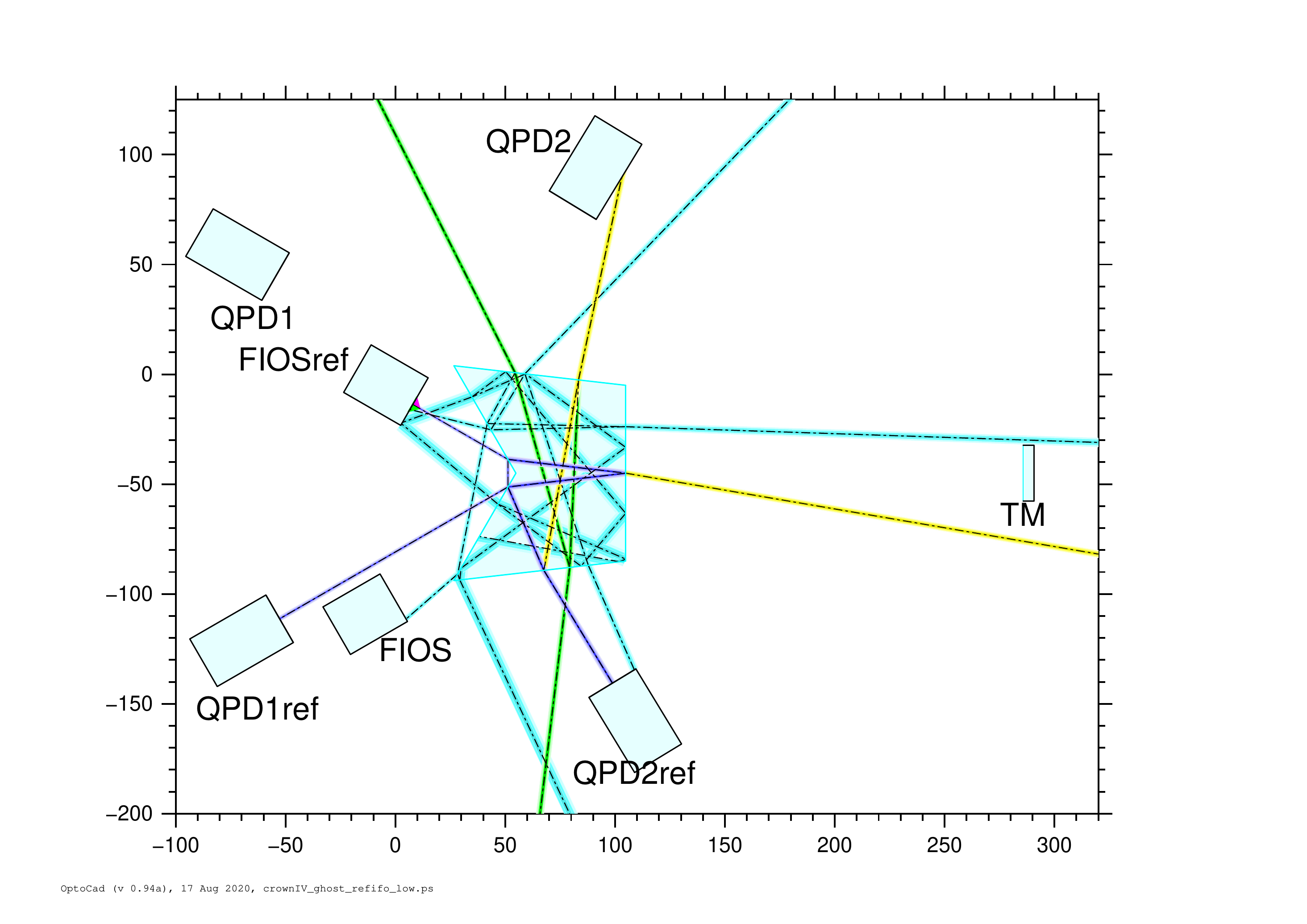}
};
\draw (-2,2) node {(b)};
\end{tikzpicture}
\caption{Ghost beam suppression in TM IFO (a) and Ref IFO (b). Absolute errors of $10^{-2}$ are introduced in the power transmission coefficient of each surface, and ghost beams down to a power threshold of $P_{\text{ghost}}/P_{\text{nom.}}\! >\! 10^{-12}$ relative to the nominal beam power are considered. The nominal beams in the TM IFO and Ref IFO are drawn in red and blue respectively. Ghost beams with $1 \!>\! P_{\text{ghost}}/P_{\text{nom.}} \!>\! 10^{-3}$ (yellow), $10^{-3} \!>\!P_{\text{ghost}}/P_{\text{nom.}} \!>\! 10^{-7}$ (green) and $10^{-7} \!>\! P_{\text{ghost}}/P_{\text{nom.}} \!>\! 10^{-12}$ (cyan) relative power level are depicted.}
\label{fig:ghost_low}
\end{figure}

\begin{figure}[htb]
\centering
\includegraphics{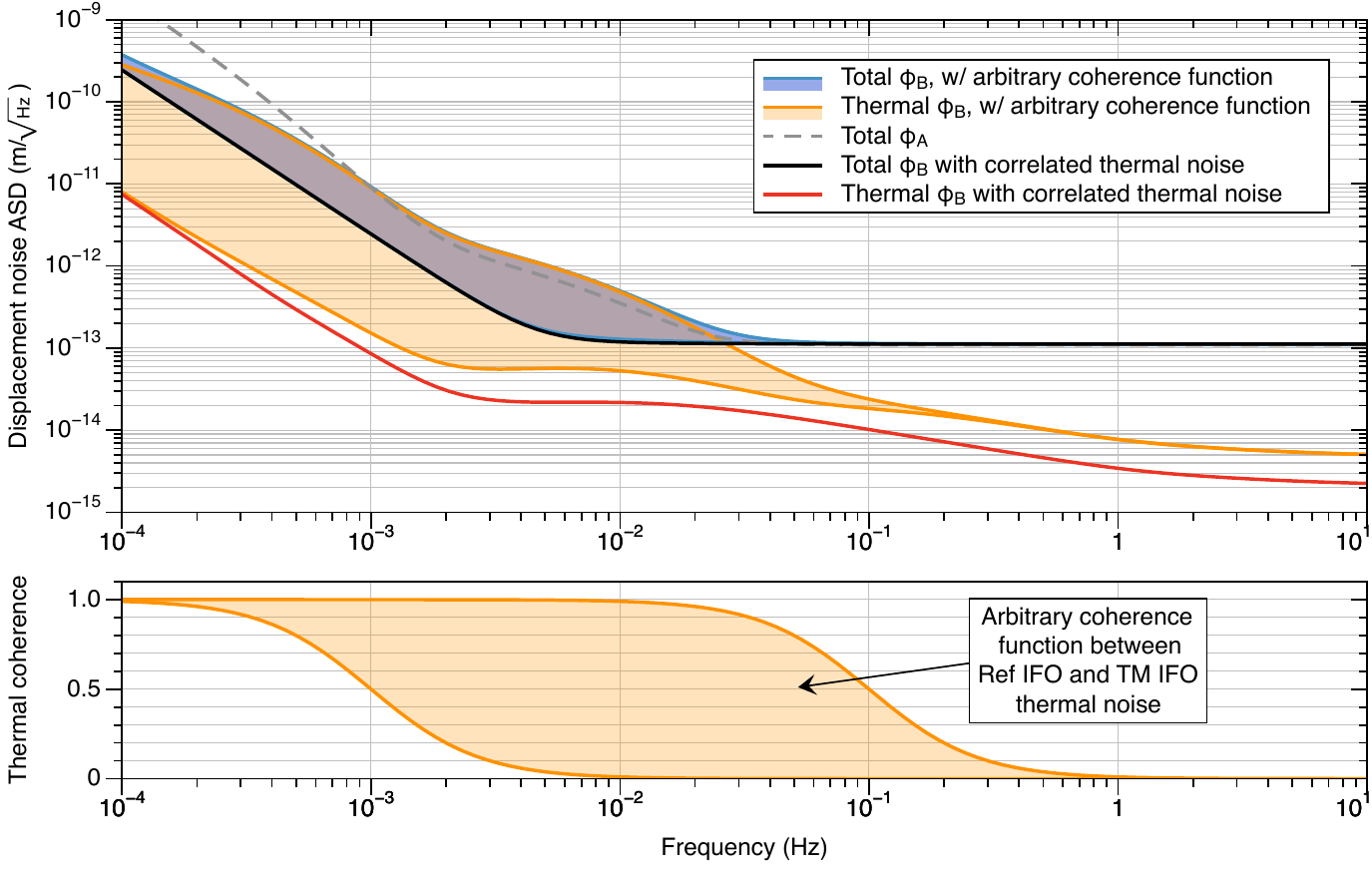}
\caption{\added{Influence of an arbitrary correlation in temperature noise between TM IFO and Ref IFO on the performance of the dual-OH setup. The coherence function is assumed to take the form $1/\sqrt{1+(f/f_{0})^2}$, where $f$ is the Fourier frequency and $f_{0}$ is the cut-off frequency, yielding perfect correlation at frequencies $f \ll f_0$, and total uncorrelation at frequencies $f \gg f_0$. In this example we show the resulting displacement noise for an arbitrarily distributed coherence function with $f_0 \in (10^{-3},10^{-1})\,$Hz.}}
\label{fig:combiB_cohTMRef}
\end{figure}

Table~\ref{tab:common_info} lists all of the remaining experiment parameters that are not included in the text. Figure~\ref{fig:ghost_low} shows ghost beams down to a power threshold of $10^{-12}$ relative to the nominal beam power. \added{Figure~\ref{fig:combiB_cohTMRef} shows the influence of an arbitrary coherence in thermal noise between the TM IFO and the Ref IFO, which was assumed to be unity in Figure~\ref{fig:Residual_displacement_noise}}.

\begin{table}[h!]
\small
\centering
\caption{\label{tab:common_info} Parameters of the whole setup. AOI is {\it Angle Of Incidence}}
\begin{tabular}{lclclclclcl}
\bhline{1.0pt}
\multicolumn{5}{c}{Optical head}                                                      \\ \hline
\multicolumn{2}{c}{(width, height, depth) {[}mm{]}}             & \multicolumn{3}{c}{(80.0, 30.0, 50.0)}             \\
\multicolumn{2}{c}{($\alpha$, $\gamma$, side angle) {[}deg{]}}               & \multicolumn{3}{c}{(120.0, 90, 6.5)}  		   \\\hline
\multicolumn{5}{c}{Quadrant photodiodes}                                                                               \\\hline
\multicolumn{2}{c}{(width, height, depth) {[}mm{]}}             & \multicolumn{3}{c}{(25.0, 25.0, 40.0)} \\
\multicolumn{2}{c}{active radius {[}mm{]}}                             & \multicolumn{3}{c}{0.25}             \\
\multicolumn{2}{c}{slit width {[}mm{]}}                         & \multicolumn{3}{c}{0.25$\times$0.05} \\
\multicolumn{2}{c}{optical distance from the prism {[}mm{]}}       & \multicolumn{3}{c}{TM:100.0, Ref: 120.0}            \\\hline
\multicolumn{5}{c}{Beams}                                                                              \\\hline
\multicolumn{2}{c}{waist radius {[}mm{]}}                       & \multicolumn{3}{c}{0.3}              \\
\multicolumn{2}{c}{waist offset from the FIOS {[}mm{]}}         & \multicolumn{3}{c}{100.0}              \\
\multicolumn{2}{c}{distance from the FIOS to the prism {[}mm{]}} & \multicolumn{3}{c}{TM:37.0, Ref:50.0}             \\
\multicolumn{2}{c}{AOI to the input surface {[}deg{]}} 		    & \multicolumn{3}{c}{60.0} 		       \\
\multicolumn{2}{c}{AOI to the TM {[}deg{]}}            		    & \multicolumn{3}{c}{9.71}			   \\\bhline{1.0pt}
\end{tabular}
\end{table}

\authorcontributions{Conceptualization, Y.Y., K.Y. and M.D.A.; methodology, Y.Y., K.Y. and M.D.A.; software, Y.Y., K.Y., C.V., T.S. and M.D.A.; validation, Y.Y., K.Y., V.H., C.V., D.P. and M.D.A.; formal analysis, Y.Y., K.Y. and M.D.A.; investigation, Y.Y., K.Y., V.H., C.V., D.P., G.F.B., T.S. and M.D.A.; writing--original draft preparation, Y.Y., K.Y. and M.D.A.; writing--review and editing, M.D.A.; visualization, Y.Y., K.Y., and M.D.A.; supervision, J.J.E.D., M.M., J.J., G.H. and M.D.A; project administration, J.J.E.D. and M.D.A.; funding acquisition, J.J.E.D. and M.D.A. All authors have read and agreed to the published version of the manuscript.}

\funding{This work has been supported by: the Chinese Academy of Sciences (CAS) and the Max Planck Society (MPG) in the framework of the LEGACY cooperation on low-frequency gravitational-wave astronomy (AEI: M.IF.A.QOP18098, MPIfR: M.IF.A.RADI 8098); Clusters of Excellence ``QuantumFrontiers: Light and Matter at the Quantum Frontier: Foundations and Applications in Metrology'' (EXC-2123, project number: 390837967); PhoenixD: ``Photonics, Optics, and Engineering – Innovation Across Disciplines'' (EXC-2122, project number: 390833453). }



\conflictsofinterest{The authors declare no conflict of interest.}

\end{document}